\documentclass[%
 reprint,
 onecolumn,
nofootinbib,
 amsmath,amssymb,
 aps,
]{revtex4-2}

\usepackage{graphicx}
\usepackage{dcolumn}
\usepackage{epsfig,latexsym,cancel,amssymb,amsmath,xcolor}
\usepackage{hyperref}         
\pdfoutput=1
\newcommand{\ben}{\begin{enumerate}}
\newcommand{\een}{\end{enumerate}}

\newcommand{\bea}{\begin{eqnarray}}
\newcommand{\eea}{\end{eqnarray}}
\newcommand{\be}{\begin{equation}}
\def\bel#1{\begin{equation} \label{#1}}
\newcommand{\ee}{\end{equation}}
\newcommand{\bi}{\begin{itemize}}
\newcommand{\ei}{\end{itemize}}
\newcommand{\ba}{\begin{align}}
\newcommand{\ea}{\end{align}}

\def\be{\begin{equation}}
\def\ee{\end{equation}}
\def\bea{\begin{eqnarray}}
\def\eea{\end{eqnarray}}

\def\ltap{\ \raise.3ex\hbox{$<$\kern-.75em\lower1ex\hbox{$\sim$}}\ }
\def\gtap{\ \raise.3ex\hbox{$>$\kern-.75em\lower1ex\hbox{$\sim$}}\ }
\def\gl{\ \raise.5ex\hbox{$>$}\kern-.8em\lower.5ex\hbox{$<$}\ }
\def\roughly#1{\raise.3ex\hbox{$#1$\kern-.75em\lower1ex\hbox{$\sim$}}}

\newcommand{\comments}[1]{}
\usepackage{color}

\begin{document}

\title{$\alpha$-attractor inflation: Models and predictions}

\author{Sukannya Bhattacharya}
\email{sukannya.bhattacharya@unipd.it}
 \affiliation{Dipartimento di Fisica e Astronomia ``G. Galilei'', Universit\`a degli Studi di Padova, Via Marzolo 8, 35131 Padova, Italy;\\
INFN, Sezione di Padova, Via Marzolo 8, 35131 Padova, Italy }
\author{Koushik Dutta}%
 \email{koushik@iiserkol.ac.in}
\affiliation{Department of Physical Sciences, Indian Institute of Science Education and Research Kolkata, Mohanpur, WB741246, India.
}%
\author{Mayukh R. Gangopadhyay}
\email{mayukh\_ccsp@sgtuniversity.org}
\affiliation{Centre For Cosmology and Science Popularization (CCSP), SGT University, Gurugram, Delhi- NCR, Haryana- 122505, India}
\author{ Anshuman Maharana}
\email{anshumanmaharana@hri.res.in}
\affiliation{Harish Chandra Research Institute, HBNI, Chattnag Road, Jhunsi, Allahabad -  211019, India}


\begin{abstract}
The $\alpha$-attractor models are some of the most interesting models of inflation from the point of view
 of upcoming observations in cosmology and also attractive from the point of view of supergravity. We confront representative models of exponential and
 polynomial $\alpha$-attractors with the latest  cosmological data (Planck'18+BICEP2/Keck array) to obtain predictions and best fit values  of model parameters.
 The analysis is done by making use of ModeChord and CosmoMC plugged together via PolyChord.
 \end{abstract}
\maketitle

\section{Introduction}
\label{Seckmi}

Cosmic inflation is the leading candidate for producing super-Hubble coherent perturbations that seed the cosmic structures and are imprinted on the cosmic microwave background  (CMB). On the other hand, 
$\Lambda$CDM cosmology  is on firm footing for describing the late-time cosmological history of the Universe. Other than the dark energy, the model includes the existence of non-baryonic dark matter and several constituents of the Standard Model. In the usual set-up of inflation, the universe is driven to exponential expansion due to the dominance of energy density of a scalar field dubbed as the inflaton. At the end of inflation, the energy density stored in the field is converted (either perturbatively or non-perturbatively) to the energy of all the constituents required by the $\Lambda$CDM model. %

While constructing and analysing a model of inflation, a few points are of immense importance in contemporary cosmology: \\
(i) The nature of the inflaton: It is desirable that the field is a natural outcome of high-energy physics  playing a role at that scale.\\
(ii) The nature of (p)rehearing: converting the energy of the inflaton to the constituents of the low energy physics requires a deeper understanding of the couplings between these fields\\
(iii) Precision analysis: each inflation model must be confronted with high-precision data to constrain model parameters. 

Over the years,  CMB  observations have become the main probe for the physics of the early Universe. In the context of inflation, the 
 fluctuations are parameterised by the strength of the power spectrum $A_s$, the scalar spectral tilt $n_s$, the tensor-to-scalar ratio $r$, and, in certain cases, the non-Gaussianity parameter $f_{\rm NL}$. All these observables encode the nature of primordial fluctuations produced by the inflaton.  
Theoretical computations needed to obtain the predictions for these observables involve the computation of the primordial fluctuations at the time when the CMB modes exit the horizon. 
In cases where the slow roll approximation is valid, consistent approximations for $n_s$ and
$r$  can be obtained by  computing the slow roll parameters at a fiducial value of the e-fold, $N_e$, defined as the number of e-folds between horizon exit
of the pivot CMB mode and the end of inflation.
Observational constraints on $n_s$ from CMB motivates to choose $N_e$ between 50 and 60. 

However, as we are deep into the era of precision cosmology, it has been widely advocated~\cite{Jm1, modecode, M1, M2, M3,Dai:2014jja,  Cicoli:2016olq} that a more rigorous numerical analysis is needed to determine model predictions.
In this type of precision analysis, it is ideal to vary both the model parameters and $N_e$ while confronting the models with data (e.g. temperature or polarization data from CMB). Furthermore the primordial fluctuations are determined by numerically solving perturbation equations, without relying
on the slow roll approximations. This is particularly important in certain  cases where it is not possible to obtain the exact form of the slow roll parameters or if the slow roll approximations are unreliable. The above works (and many more in the literature) analyse specific models but they also advocate that all models should be analysed using similar methods. In this article we will carry out such an analysis for 
a class of models which are now being considered as particularly important for future observations : the $\alpha$ attractors, which we describe next.
The main goal of this work is to mark the difference in the numerically obtained best-fit values of the parameters for the two main classes of $\alpha$-attractor models: exponential $\alpha$ attractors and polynomial $\alpha$ attractors.

        A key prediction of the inflationary model is the presence of primordial tensor fluctuations. The 
        ratio of tensor and scalar power spectra represented by $r$ is related to the slow roll parameter $\epsilon$ at the time
of horizon exit of the modes of interest. Similarly, the scalar spectral tilt $n_s$ is determined by a linear combination of $\epsilon$ and $\eta$. From the observational point of view, a red spectral
tilt is well established, yet there is no observation of a tensor component in the fluctuations. This implies that inflationary models consistent with data have to be special in the sense
that there is a hierarchy in the magnitude of the two slow-roll parameters so that they correctly predict the observed tilt and are consistent with the lack of observational evidence for $r$ so far (i.e., upper bound on $r$).
Motivated by this, starting from nearly a decade ago\footnote{Models proposed even earlier are consistent with
the present data see, e.g., the Starobinsky model \cite{Starobinsky:1980te},  the GL model \cite{Goncharov:1983mw,Goncharov:1984jlb,Linde:2014hfa},  the Higgs inflation \cite{Futamase:1987ua,Salopek:1988qh,Bezrukov:2007ep}}, a large family of different inflationary theories predicting the same characteristic form of observables were proposed and studied in  \cite{Kallosh:2013hoa,Kallosh:2013pby,Kallosh:2013lkr,Ferrara:2013rsa, Kallosh:2013maa, Kallosh:2013tua, Ellis:2013xoa, Buchmuller:2013zfa,  Kallosh:2013yoa, Galante:2014ifa,Kallosh:2015zsa,Kumar:2015mfa,Kallosh:2019eeu,Kallosh:2019hzo, Ferrara:2016fwe,Kallosh:2017ced,Gunaydin:2020ric,Kallosh:2021vcf, LiteBIRD:2022cnt, Kallosh:2022feu, Odintsov:2022bpg, Odintsov:2020thl, Odintsov:2016vzz}. From the point of view of observations, these models have the attractive feature that the model parameters
allow for a regime in which $n_{s} = \mathcal{O}(1/N_{e})$ while $r = \mathcal{O}(1/N_{e}^2)$ (where $N_{e}$ is the number of e-folds between horizon exit of CMB modes and the end of inflation). Another appealing
feature of these models is that a large class have been successfully embedded in supergravity. In view of this,
these models have been proposed as benchmarks for future cosmological experiments\footnote{Such as  CMB stage 4, LiteBIRD and CORE~\cite{Aba}.} (see, e.g., \cite{LiteBIRD:2022cnt}). The models are collectively known as $\alpha$-attractors, where $\alpha$ is a parameter that affects the observables.

  Given the importance of these models in the realm of contemporary precision cosmology, they certainly deserve detailed studies. The goal of this work is to initiate a systematic cosmological analysis of these models as is needed to confront them with current data.
 As mentioned earlier, this is in the spirit of the studies  of \cite{Jm1, modecode, M1, M2, M3, Dai:2014jja, Cicoli:2016olq}. Further studies in this direction have been carried out in  \cite{kcmb, Bhattacharya:2020gnk, Cicoli:2020bao, Neves:2020anh}. Here, we will  follow \cite{modecode}, making use of
CosmoMC \cite{Lewis:2002ah} and ModeChord\footnote{Publicly available at http://modecode.org/modechord .} plugged together through Polychord \cite{Handley:2015fda}. 

Analysing an inflation model using ModeChord has two advantages.
Firstly, ModeChord analyses inflationary dynamics without relying on the slow roll approximation; thus slow-roll violating effects
which can be important for confronting models with precision data are automatically captured. Also, in cases where the exact analytic form
of the slow roll parameters is not available or available only under some approximation it is natural and desirable to adopt this method. 
The second advantage is related to the reheating epoch.
Given a model of inflation, the  precise predictions for various cosmological observables
(such as the spectral tilt) are sensitive to the post inflationary history of the model - this is an input for determining $N_e$ which thereby sets the value of the observables such as the spectral tilt. In the post-inflationary history, most of our ignorance lies in the reheating epoch. ModeChord allows to incorporate
this ignorance into the analysis systematically, by considering the so called ``generalised reheating scenario''. Here, the duration of the reheating epoch is treated as a parameter and is sampled over.

  This paper is structured as follows. In Sec.~I \ref{models}, we briefly review $\alpha$-attractor models. In Sec.~\ref{analysis}, we describe our methodology
and present our results. There are a large number of $\alpha$-attractor models, to keep the paper streamlined we will focus on some representative
models: we will consider the $T$-model from the ``exponential $\alpha$-attractors'' and the models of first and second kind from the ``polynomial $\alpha$-attractors''. 
Our choice of models is guided by the following: as emphasised in \cite{Kallosh:2022feu}, the exponential T-models and polynomial models  nicely complement each other; in the $n_s-r$ plane, the
exponential models cover the left half and the polynomial models lie to the right 
spanning the entire interesting region 
observationally allowed by the Planck/BICEP/Keck data (see Fig. 1 of ~\cite{Kallosh:2022feu}). We discuss and conclude in Sec.~\ref{disc}.

\section{$\alpha$-attractor models}

\label{models}
A large class of inflationary models (e.g the Starobinsky model, chaotic inflation model  $\lambda \phi^{4} $ with non-minimal coupling to gravity and
conformal, superconformal and supergravity generlisations \cite{Starobinsky:1980te, Salopek:1988qh,Linde:2011nh,Kaiser:2013sna, Kallosh:2013pby,Kallosh:2013lkr,Kallosh:2013hoa,Ferrara:2013rsa,Kallosh:2013maa,Kallosh:2013tua,Ellis:2013xoa,Buchmuller:2013zfa}) have a common prediction for  the scalar spectral index $(n_s)$ and the tensor-to-scalar ratio $(r)$:
\bea
n_s &= 1-\frac{2}{N_e}\nonumber \\
r &= \frac{12}{N_e^2},\label{nsr_Star}
\eea
 in the limit of large $N_e$, where $N_e$ is the number of e-folds of inflation, counted from the end of inflation to the horizon exit of the CMB modes. 
 
 These models can be generalized with the introduction of a new parameter $\alpha$, (which is inversely related to the curvature in the 
 field space of the inflaton). This class of attractor models of inflation, known as $\alpha$-attractors, also lead to a general prediction for inflationary observables~\cite{Kallosh:2013yoa}:
 \bea
n_s &= 1-\frac{2}{N_e}\nonumber \\
r &= \frac{12\alpha ^2}{N_e^2}.\label{nsr_alp}
\eea
 Popular examples of models of this kind are Starobinsky inflation~\cite{Starobinsky:1980te}, where the Jordan frame action contains polynomials of the Ricci scalar (leads to $\alpha =1$), Higgs inflation~\cite{Futamase:1987ua,Salopek:1988qh,Bezrukov:2007ep} with nonminimal gravitational couplings $g(\phi , R)$, GL model~\cite{Goncharov:1983mw,Goncharov:1984jlb,Linde:2014hfa}, etc. For these models the Einstein frame potential can be expressed as
 \be 
V(\phi) = V_0\bigg(1-e ^{(-\frac{\phi}{\sqrt{6}\alpha})}\bigg)^2 .
\label{Valp}
\ee
$\phi$ is the canonical field related to the field $\rho$ as
\be
\rho =e ^{(-\frac{\phi}{\sqrt{6}\alpha})},
\label{rhophi}
\ee
 and $\rho$ defines the potential $V_J(\rho)$ in the Jordan frame.
$\alpha$ can be much smaller than 1, and this leads to very small values of $r$. This is interesting in the context of the upcoming experimental efforts with CMB B-modes to probe the tensor fluctuations of inflation (see, e.g., \cite{Kallosh:2019hzo}). On the other hand, large values of $\alpha$ render the predictions to be the same as chaotic inflation $V(\phi) = \frac{1}{2}m^2\phi ^2$. Note that these models are ``exponential" in the sense that the
inflaton plateau is approached exponentially in $\phi$.  

A class of models that can be easily embedded in supergravity are examples which lead to similar predictions for inflationary observables as in Eq.~\eqref{nsr_alp} upto first order in $1/N_e$.
For the generalized ``T-models", 
\cite{Kallosh:2013yoa} the inflaton potential takes the form:
\bea
V(\phi) &=V_0 \tanh ^{2n} (\frac{\phi}{\sqrt{6}\alpha}).
\label{tanh2}
\eea
%
The full predictions for these models are
\bea
n_s &= \frac{1-\frac{2}{N_e}-\frac{3\alpha ^2}{4N_e^2}+\frac{1}{2nN_e}(1-\frac{1}{N_e})g(\alpha , n)}{1+\frac{1}{2nN_e}g(\alpha ,n)+\frac{3\alpha ^2}{4N_e^2}}\nonumber \\
r &= \frac{12\alpha ^2}{N_e^2(1+\frac{1}{2nN_e}g(\alpha ,n)+\frac{3\alpha ^2}{4N_e^2})}.
\label{nsrtanhgen}
\eea
where $g(\alpha ,n)=\sqrt{3\alpha ^2(4n^2+3\alpha ^2)}$.  This shall be the first model that we will analyse in detail\footnote{As mentioned
in the introduction, our analysis incorporates ModeChord, hence, it will not rely on the expression~\ref{nsrtanhgen}, which are provided to give the reader an idea
of the nature of the model.}. Note that in the limit of large $N_e$, Eq.~\eqref{nsrtanhgen} reduces to the form in Eq.~\eqref{nsr_alp}.

  Another class of $\alpha$-attractors are the so called ``polynomial" $\alpha$-attractors \cite{Kallosh:2022feu}
 (these have their origins in various models of inflation such as brane and pole inflation, e.g.,  \cite{Kallosh:2018zsi, Kallosh:2019hzo, Dvali:1998pa, Dvali:2001fw, Burgess:2001fx,Kachru:2003sx,Lorenz:2007ze,Martin:2013tda, Kallosh:2018zsi, Stewart:1994pt, Fairbairn:2003yx, Dong:2010in, dimo,yogesh, Galante:2014ifa, Terada:2016nqg, Karamitsos:2019vor, Kallosh:2019hzo, Kallosh:2021mnu}). Here, instead of an exponential approach, the 
  inflationary plateau is reached with an inverse power of the field :  $V \sim V_0 [1-(\frac{\mu}{\phi})^{k} + ...]$.  
  The simplest polynomial $\alpha$-attractor models have a potential (for the canonically normalised
  field which has a relation with $\rho$ same as Eq.~\eqref{rhophi})
\be
V(\phi) = V_0\frac{\phi ^{2n}}{\phi ^{2n} + \mu ^{2n}}.\label{Vlog1can}
\ee
The potential for the canonically normalized $\phi$ is similar to the class of models known as power law plateau potential which has supergravity origin (for more details the author is advised to consult~\cite{dimo}and \cite{yogesh}).
In the slow roll approximation and limit of large $N_e$,
\begin{eqnarray}
n_s & = & 1-\frac{2n+1}{n+1}\frac{1}{N_e},\nonumber \\
r &=& 2^{\frac{3+n}{1+n}}\bigg(\mu ^{2n}n\bigg)^{\frac{1}{1+n}}\bigg( (n+1)N_e\bigg)^{-\frac{1+2n}{1+n}}.
\label{gen_nsr_polyattr}
\end{eqnarray}
These expressions are valid in the approximation $\mu \ll 1$. Our analysis with ModeChord will 
be numerical, and hence will not rely on this approximation. The analysis will also
involve varying $\mu$ as a parameter.
For large $n$, the above approach the form of $n_s$ and $r$ in Eq.~\eqref{nsr_alp} with
 $\alpha ^2 = \frac{1}{6}\bigg(\frac{\mu}{n^2}\bigg)^2$. The polynomial $\alpha$-attractor models of the second kind have the following form in the Einstein frame
\be
V(\phi) = V_0\frac{(\phi ^2 +\mu ^2)^{k/2}-\mu ^k}{(\phi ^2 +\mu ^2)^{k/2}+\mu ^k}.\label{Vlog2can}
\ee
Again, it is easy to see that the inflationary plateau is approached as a power law in $\phi$. The analytic forms of
$n_s$ and $r$ are available (again in the case of $\mu \ll 1$), but they are rather complicated in the general case.
 \section{Analysis}
  \label{analysis}
Now, we consider the analysis of the following models:\\
(i) $n=1$ $\tanh$ model given by Eq.~\eqref{tanh2},\\ 
(ii) $n=2$ polynomial of the first kind given by Eq.~\eqref{Vlog1can}, \\ 
(iii) and $k=1$ polynomial of the second kind given by Eq.~\eqref{Vlog2can}.\\
For the case (i), $\alpha$ is the only remaining model parameter to be varied, whereas for the cases (ii) and (iii), $\mu$ is the model parameter that is varied. While the predictions for the model (i) are expected to conform to the $\alpha$-dependent forms of $n_s$ and $r$ in Eq.~\eqref{nsr_alp}, those for models (ii) and (iii) are expected to have the $n$ and $\mu$-dependent form in Eq.~\eqref{gen_nsr_polyattr}. Therefore, studying only these three models actually covers both the sectors of exponential $\alpha$-attractors and polynomial $\alpha$-attractors, and this is the reason for choosing and studying these three models only. 
\begin{figure}[]
\centering
\includegraphics[width=0.95\textwidth]{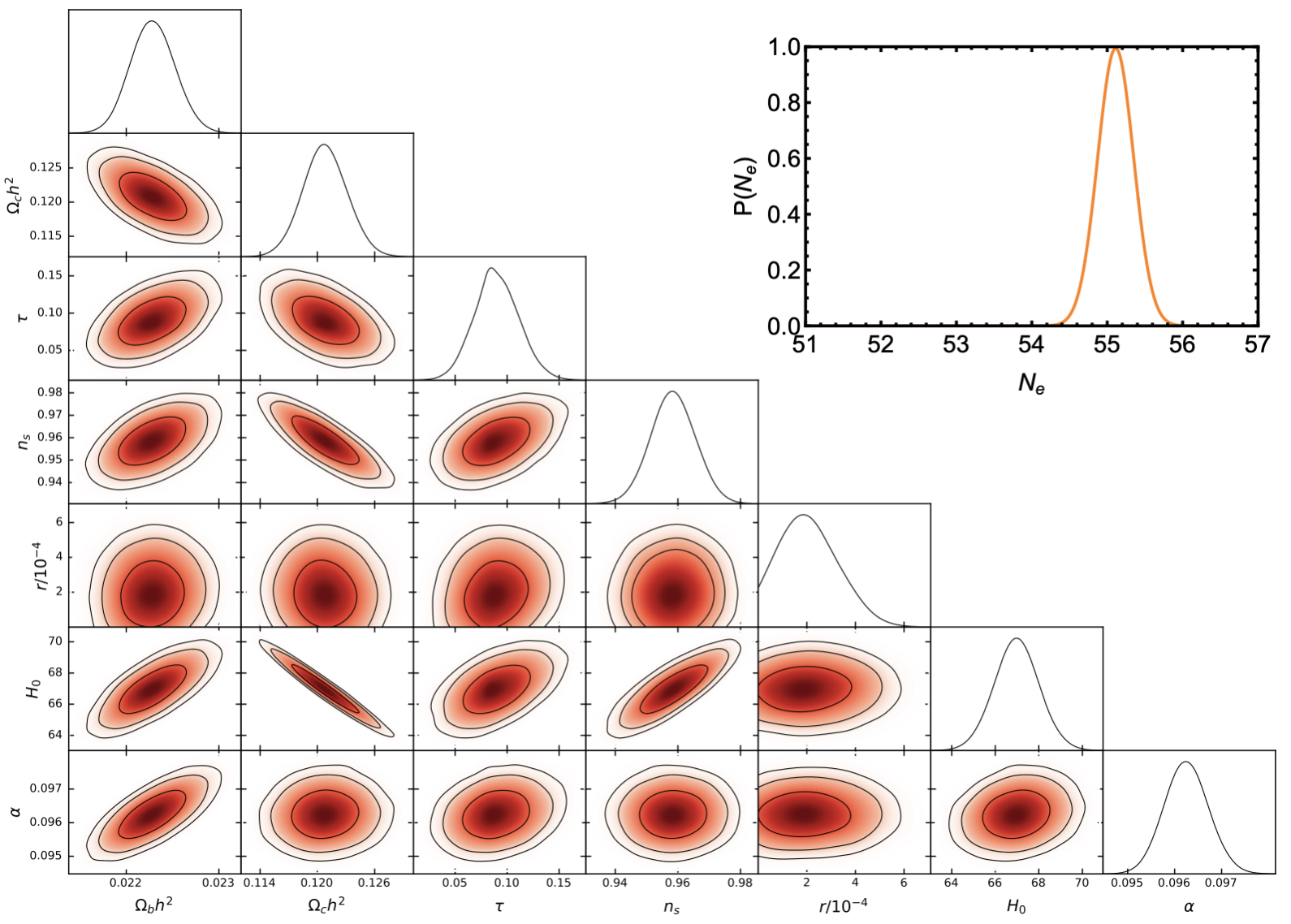}
\caption{One dimensional posterior distribution and two dimensional joint contours at $68\%$, $95\%$ and $99\%$ confidence limit (C.L.) for parameters. \textbf{Inset:}  One dimensional posterior distribution for number of e-folds($N_e$) for n=1 the T- model given by Eq.~\eqref{tanh2}. }
\label{tanhn1plots}
\end{figure}

The analysis is carried out using publicly available packages: CosmoMC and ModeChord brought together through Polychord. Given a  inflationary potential, ModeChord numerically computes the primordial scalar and tensor power spectra by directly solving the perturbation equations. These primordial spectra are then put through CAMB \cite{camb}
 in the CosmoMC package
with the help of the plug-in package Polychord and then evolved through transfer functions. Then the angular power spectra obtained at CMB from the theory are compared to 
the observed fluctuations (e.g. by Planck) using CosmoMC.  In the usual implementation of ModeChord, the cosmological perturbations are evaluated without assuming slow-roll conditions, therefore, our analysis does not make use of the expressions for $n_s$ and $r$ in Eqs~\eqref{nsrtanhgen} and~\eqref{gen_nsr_polyattr}. The best-fit values of the potential parameters are estimated using CosmoMC without relying on the slow-roll approximation. 
In this analysis, flat priors are used in appropriate ranges for all the model parameters.
\begin{figure}[]
\centering
\includegraphics[width=0.95\textwidth]{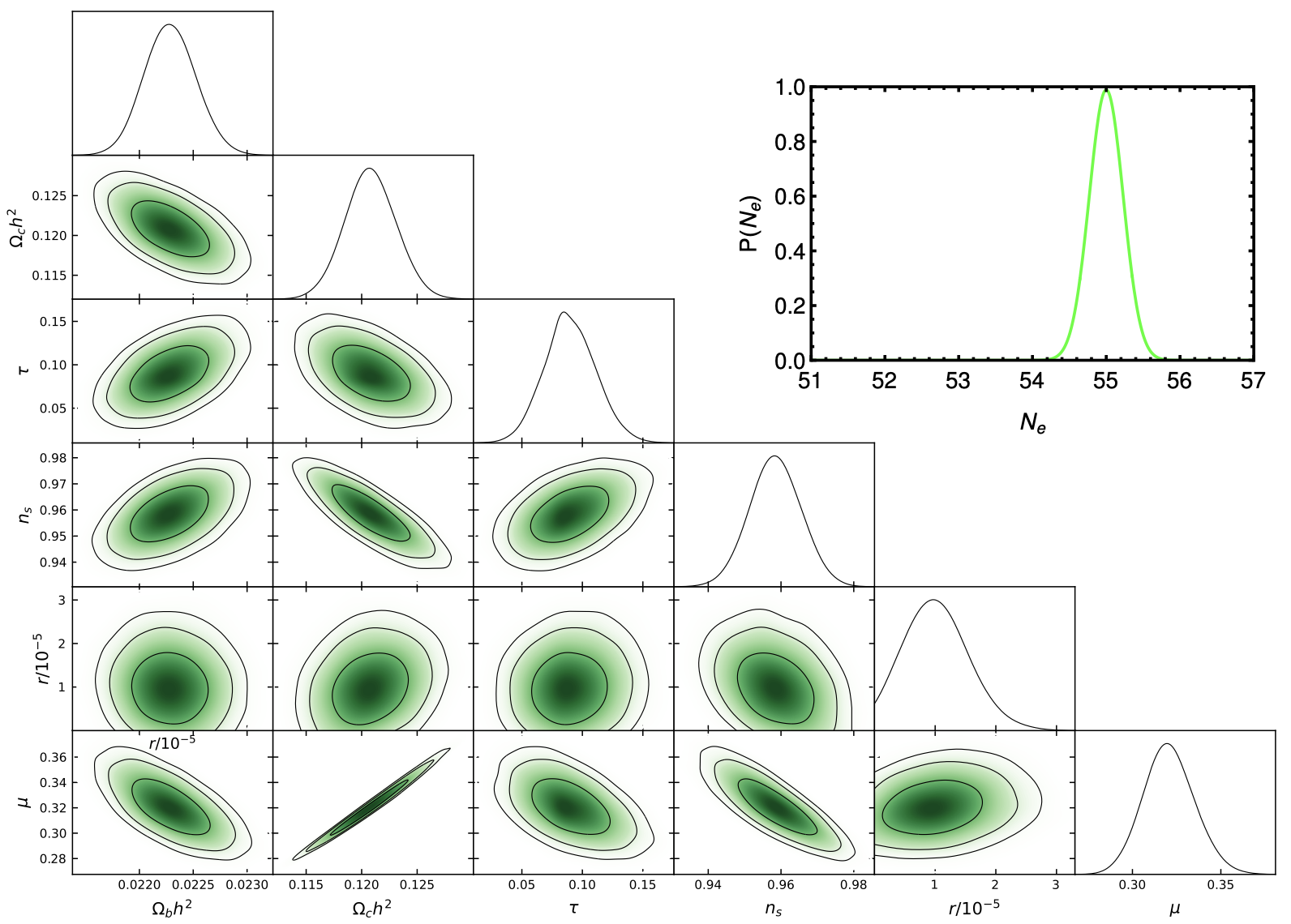}
\caption{One dimensional posterior distribution and two dimensional joint contours at $68\%$, $95\%$ and $99\%$ C.L. for parameters. \textbf{Inset:} One dimensional posterior distribution for number of e-folds($N_e$) for n=2 of the polynomial attaractors of first kind given by Eq.~\eqref{Vlog1can}. }
\label{plotsan2}
\end{figure}

The simplest check for a cosmological model can be done in the base 6-dimensional parameter space described by: the baryon and cold dark matter energy densities ($\Omega_{b}h^2$, $\Omega_{c}h^2$), the acoustic peak angular scale($\theta$), the optical depth to reionization ($\tau$), the amplitude ($A_s$) and the spectral index ($n_s$) of the primordial power spectrum. 
For given values of the model parameters, we solved for the cosmological perturbations by solving perturbation equations using ModeChord (together with CosmoMC, through Polychord) without the assumption of slow-roll. The Boltzmann solver CAMB is used to evaluate the two-point correlation functions for temperature and polarization anisotropies. Then, the model parameters are estimated and the goodness of the fit is determined using CosmoMC by comparing the model predictions with CMB data. In addition, the number of e-folds of inflation ($N_e$) has also been set as a variable. Therefore, in this work, we have varied the four late-time cosmological parameters $\Omega_{b}h^2$, $\Omega_{c}h^2$, $\theta$ and $\tau$, the model parameters from inflation and $N_e$. The primordial power spectrum is calculated by the code at different points of the parameter space of the inflation model parameters and $N_e$. The constrains these parameters as well as $n_s$ and tensor-to-scalar ratio $r$ are provided. The amplitude $A_s$ can be found as a function of the model parameter $V_0$ (see Eq.~\eqref{scalarstrength}). Using this approach confirms the removal of the otherwise uncertainty from the analysis of the inflationary parameter space and allows us to constrain inflationary models with complicated spectra not well described otherwise.

The connection between
inflationary dynamics and primordial fluctuations requires $N_e$: the number of e-foldings between horizon exit and the end of inflation. This is sensitive to
the post-inflationary history of the universe and hence affected by the uncertainties associated with reheating. ModeChord deals with this  uncertainty by
working with the generalised reheating scenario (GRH) -  $N_e$ is varied over a physically well motivated range and optimised over. If the 
universe under goes $N_{\rm re}$ e-foldings during the reheating epoch and has an  effective equation of state $w_{\rm re}$ during the epoch then $N_e$ is given by
\be
N_e = N_{\rm IRH}- \frac{1}{4} (1- 3 w_{re}) N_{\rm re},
\label{npnirh}
\ee
where $N_{\rm IRH}$ is the value of $N_e$ assuming instantaneous reheating. Various physical arguments can be given for $w_{\rm re} < 1/3$
(see, e.g.,~\cite{Dai:2014jja,Allahverdi:2020bys}  for a discussion and further references). Given this, Modechord takes $N_{\rm IRH}$ as an upper bound
for the value of $N_e$ and varies it in the range  $N_{\rm IRH} > N_e  > 20$. The lower limit comes from
the requirement that at the end of inflation, all the relevant cosmological scales are well outside of the horizon.

\begin{figure}[]
\centering
\includegraphics[width=0.95\textwidth]{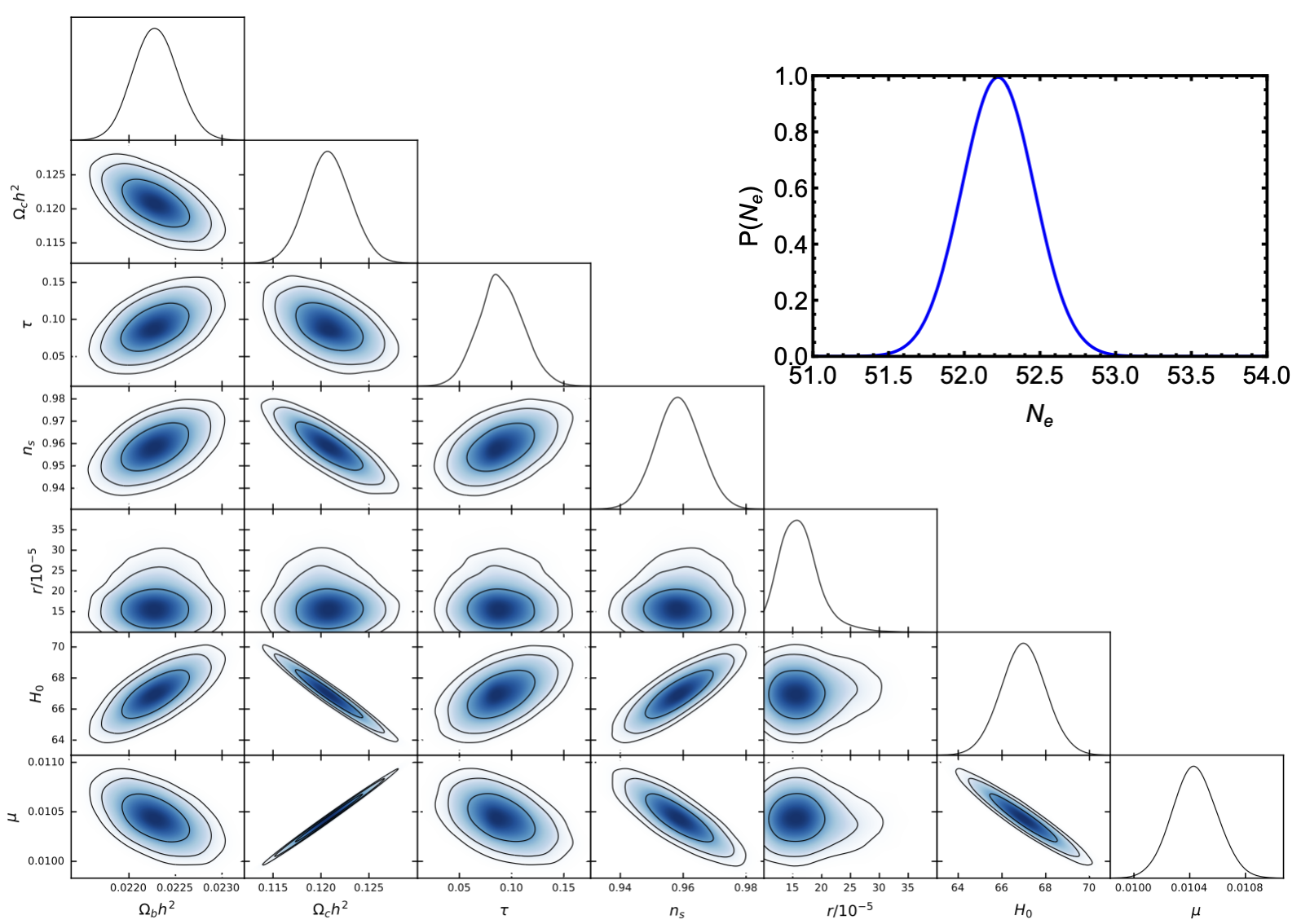}
\caption{One dimensional posterior distribution and two dimensional joint contours at $68\%$, $95\%$ and $99\%$ CL for parameters. \textbf{Inset:} One dimensional posterior distribution for number of e-folds($N_e$) for k=1 of the polynomial potential of second kind given by Eq.~\ref{Vlog2can}. }
\label{plotscase2k1}
\end{figure}

The value of $N_{\rm IRH}$ can be obtained by tracking the
energy density of the universe from the end of inflation to today~\cite{modecode,Adshead:2010mc}:
\begin{equation}
\label{matchingeq}
  N_{\rm IRH} =  55.75 -  \log \left[ { 10^{16} \rm{Gev} \over  V_{\rm end}^{1/4} }\right].
\end{equation}
Furthermore, up to logarithmic corrections one can replace $V_{\rm end}$ by the  plateau value $(V_0)$ in the above expression. The expression
for the strength of the scalar perturbations
  \begin{equation}
  \label{scalarstrength}
  A_s \approx { 2 \over {3 \pi^{2} r} } \left( { V_{0} \over M^{4}_{\rm pl} } \right)
  \end{equation}
can then be used to express $N_{\rm IRH}$ in terms of  $A_s$ and $r$.

  We present our results in the form of tables and plots. The likelihoods used  are from $Planck~2018~TT+TE+EE + lowP + lensing$ and $Planck+BICEP2/Keck$ array joint analysis \cite{datacombined}.  Fig.~\ref{tanhn1plots},~\ref{plotsan2} and~\ref{plotscase2k1} show the triangle plots for the models T-model with $n=1$ (Eq.~\eqref{tanh2}),  the $n=2$ polynomial $\alpha$-attractors of the first kind  (the potential in Eq.~\eqref{Vlog1can} ) and for the $k=1$ polynomial $\alpha$-attractors of the second kind  (the potential in~\eqref{Vlog2can}) respectively.\footnote{The convergence of all the simulations is ensured satisfying the Gellman-Rubin criterion. In the case of the three analysis presented in Fig.s~\ref{tanhn1plots},~\ref{plotsan2} and~\ref{plotscase2k1} , the convergences attained are $R-1 \sim 0.0004,~0.0210,~0.1039$ respectively.}
  
  Triangle plots are shown for the marginalised posterior distributions of the inflationary and base $\Lambda$CDM parameters. For each case, the posterior distribution of the number of inflation e-folds is also plotted. Table~\ref{tbl_3model} summarises the best-fit values for the parameters varied in these three cases.
\begin{center}
\begin{table}[htb!]
\begin{tabular}{|l|l|l|l|}
\hline
Parameter        & T-model with $n=1$ & polynomial $\alpha$-attractor  &  polynomial $\alpha$-attractor                     \\ 

       &    &(1st kind) with $n=2$ &(2nd kind) with $k=1$                      \\ \hline
\hline
$\Omega_b h^2$ & $0.02227\pm{0.00023}$  & $0.02228\pm 0.00024$ & $0.02230\pm 0.00025$           \\ 
\hline

$\Omega_c h^2$ & $0.12110\pm{0.00235}$  & $0.12081\pm 0.00225$ & $0.11990\pm 0.00220$        \\ 
\hline

$\tau$ & $0.08981_{-0.02116}^{+0.02141}$  & $0.08080_{-0.01931}^{+0.01952}$ & $00.08065_{-0.01847}^{+0.01855}$         \\ 
\hline
$H_0$            & $66.9060\pm{1.0985}$ & $66.9801\pm{1.02921}$  & $66.8705\pm{1.1105}$        \\ \hline
$n_s$            & $0.96450^{+0.0068}_{-0.0073}$ & $0.96750^{+0.0071}_{-0.0070}$ & $0.97147^{+0.0002}_{-0.0002}$ \\ \hline
$r/10^{-5}$         & $2.1290^{+0.9429}_{-2.0693}$ & $18.7001^{+4.5852}_{-6.0335}$ &
$16.2980^{+2.3986}_{-3.9336}$\\ \hline
$\alpha$            & $0.0962^{+0.00046}_{-0.00047}$ & -- &   --   \\ \hline
$\mu$            & --
 & $0.31075^{+0.013633}_{-0.015222}$  &  $0.01043^{+0.0001916}_{-0.0001265}$     \\ \hline
$N_e$            & $55.1121\pm{0.1902}$ & $55.1241\pm 0.2306$ & $52.2232\pm 0.3000$             \\ \hline
$V_0 /10^{15}$   & $1.9598\pm{0.25345}$ & $3.3787\pm{1.38700}$ & $3.2599\pm{1.02855}$           \\ \hline
\end{tabular}
\caption{Observational constraints at $68 \%$ CL  on both independent and derived cosmological parameters for the T- model with $n=1$, Polynomial $\alpha$ attractor of the first kind with $n=2$ and Polynomial $\alpha$ attractor of the second kind with $k=1$
using the following data combination: $Planck~2018~TT+TE+EE + lowP + lensing$ and $Planck+BICEP2/Keck$ array joint analysis \cite{datacombined}.}
\label{tbl_3model}
\end{table}
\end{center}
\section{Results and Discussions}
\label{disc}
  The $\alpha$-attractor models are very interesting from the point of view of upcoming observations in cosmology. Motivated by this,
  we confronted them with the latest cosmological data and obtained the best-fit values of model parameters. The distributions
  of $\alpha$ and $\mu$ suggest that they would have to be generated from a high scale.  In the 2-dimensional marginalised posterior distribution in the $n_s-r$ plane, our precision analysis shows that
exponential models cover the left half and the polynomial models lie to the right 
spanning the entire interesting region. Therefore, predictions for $n_s$ are higher in polynomial $\alpha$-attractor models than those for the exponential $\alpha$-attractor models -- confirming earlier expectations~\cite{Kallosh:2022feu}. 
The best-fit values of $r$ are low, and the polynomial models typically predict $r$ values higher by one order of magnitude than the T-model, which can be seen in Table~\ref{tbl_3model}. It is interesting to note that, $r$ is severely constrained at the level of $\mathcal{O}(10^{-5})$. This shows the dependence of $r$ on $\mu$. To keep all the cosmological observables such as $A_s$ and $n_s$ in the allowed range, $\mu$ gets severely constrained. This constraint on $\mu$ propagates to $r$. Moreover, the polynomial model of the second kind requires less number of e-folds of inflation compared to the other two examples we have analysed. As mentioned earlier, the models that are not studied in this paper are equivalent in their predictions of $n_s$ and $r$ to one of the three models already analysed in the paper, since the studied cases cover both the exponential and polynomial types of $\alpha$-attractors.


Our result
  should be useful for various directions. One can now think about what embeddings of these models in string/M-theory can lead
  to the parameters in the desired range. Recently, it has been found that
  models  closely related to those analysed 
  in this paper (the hybrid $\alpha$-attractors \cite{Kallosh:2022ggf}) have a rich phenomenology in the context of  primordial black holes and gravitational waves  \cite{Braglia:2022phb,Fu:2022ypp}. It will
  be interesting to carry an analysis similar in spirit to this paper and obtain the implications for their phenomenology. We hope to return to some of these questions.\\
  \\
\\
\\
\textbf{Acknowledgements:}\\
S.B. is supported by the ``Progetto di Eccellenza'' of the Department of Physics and Astronomy of the University of Padua. She also acknowledges support by Istituto Nazionale di Fisica Nucleare (INFN) through the Theoretical Astroparticle Physics (TAsP) project. The work of K.D. is partially supported by the Indo-Russian grant No. DST/INT/RUS/RSF/P-21, No. MTR/2019/000395, and Core Research Grant No. CRG/2020/004347 funded by SERB, DST, Government of India.
 Work of M.R.G. is supported
by DST, Government of India under the Grant Agreement No. IF18-PH-228 and by Science and Engineering Research Board (SERB), DST, Government of India under the Grant Agreement No. CRG/2022/004120 (Core Research Grant). A.M. is supported
in part by the SERB, DST, Government of India by the grant No. MTR/2019/000267.

\end{document}